\documentclass[preprint,prd,aps,tighten,nofootinbib,amssymb]{revtex4}

\usepackage{graphicx}
\usepackage{epsfig,amssymb,amsmath}

\pdfoutput=1
\usepackage{subfigure}
\usepackage[colorlinks=true,linkcolor=blue,citecolor=blue]{hyperref}
\usepackage{float}

\makeatletter
\def\Hline{%
\noalign{\ifnum0=`}\fi\hrule \@height 2pt \futurelet
\reserved@a\@xhline}
\makeatother

\newcommand{\beq}{\begin{equation}}
\newcommand{\eeq}{\end{equation}}
\newcommand{\bea}{\begin{eqnarray}}
\newcommand{\eea}{\end{eqnarray}}
\newcommand{\bear}{\begin{array}}
\newcommand {\eear}{\end{array}}
\newcommand{\bef}{\begin{figure}}
\newcommand {\eef}{\end{figure}}
\newcommand{\bec}{\begin{center}}
\newcommand {\eec}{\end{center}}
\newcommand{\non}{\nonumber}

\def\GEV#1{10^{#1}{\rm\,GeV}}

\def\lrfp#1#2#3{ \left(\frac{#1}{#2} \right)^{#3}}

\begin{document}
\draft
\tighten
\preprint{ICRR-Report 684-2014-10, TU-972, IPMU14-0129}
\title{\large \bf
Relaxing Isocurvature Bounds on String Axion Dark Matter
%\\ by Thermal Inflation
}
\author{  
    Masahiro Kawasaki\,$^{a,c}$\footnote{email: kawasaki@icrr.u-tokyo.ac.jp},
    Naoya Kitajima\,$^{b}$\footnote{email:kitajima@tuhep.phys.tohoku.ac.jp},
    Fuminobu Takahashi\,$^{b,c}$\footnote{email: fumi@tuhep.phys.tohoku.ac.jp}
    }
\affiliation{
$^a$ Institute for Cosmic Ray Research, University of Tokyo, Kashiwa 277-8582, Japan\\
$^b$ Department of Physics, Tohoku University, Sendai 980-8578, Japan\\
$^c$ Kavli IPMU, TODIAS, University of Tokyo, Kashiwa 277-8583, Japan
}
%\date{\today}

\vspace{2cm}

\begin{abstract}
        If inflation scale is high,  light scalars acquire large quantum fluctuations during inflation.       
        If sufficiently long-lived, they will give rise to CDM isocurvature perturbations,  which are highly 
 	constrained by the Planck data. 
	Focusing on string axions as such light scalars, we show that thermal inflation can provide a sufficiently large entropy production to dilute the CDM isocurvature
	perturbations. Importantly, efficient dilution is possible 
	for the string axions,  because effectively no secondary coherent oscillations are induced at the end of thermal inflation, 
	in contrast to the moduli fields. We also study the viability of the axion dark matter with mass of about $7$\,keV
	as the  origin of the $3.5$\,keV X-ray line excess, in the presence of large entropy production. 
\end{abstract}

\maketitle

%%%%%%%%%%%%%%%%%%%%%%%%%%%%%%%%%%%%%%
\section{Introduction}
\label{sec:1}
%%%%%%%%%%%%%%%%%%%%%%%%%%%%%%%%%%%%%%
Inflation~\cite{Guth:1980zm} elegantly solves theoretical problems of the standard cosmology such as the 
horizon problem\footnote{
The exponentially expanding universe was also studied in Refs.~\cite{Brout:1977ix,
Starobinsky:1980te,Kazanas:1980tx,Sato:1980yn}.}
, and the slow-roll inflation paradigm is consistent 
with the observations including  the comic microwave background (CMB) and large-scale structure data. 
Measuring  the primordial B-mode polarization of CMB is therefore of crucial importance, as it would provide
a definitive proof of inflation~\cite{Starobinsky:1979ty}.

Recently the BICEP2 collaboration reported detection of the B-mode polarization, which could
be due to the primordial gravitational waves~\cite{Ade:2014xna}. If this is the case, the BICEP2 results can be
explained by the tensor mode perturbations with a tensor-to-scalar ratio, $r = 0.20^{+0.07}_{-0.05}$, whereas the central value (and therefore
significance for the signal) depends on models for foreground dust polarization~\cite{Flauger:2014qra}.

Taken at face value, the BICEP2 result strongly suggests high-scale inflation with the Hubble parameter,
$H_{\rm inf} \sim   \GEV{14} $ \cite{Ade:2014xna}.
Importantly, any light scalar particles acquire large quantum fluctuations of order $H_{\rm inf}/2\pi$ during inflation. 
Those scalars are copiously produced by coherent oscillations when their mass becomes comparable to the 
Hubble parameter after inflation. If some of them are long-lived, the coherent oscillations will contribute to dark matter, 
giving rise to isocurvature perturbations,  which are tightly constrained by the CMB observations~\cite{Ade:2013zuv}.

The QCD axion,  a pseudo Nambu-Goldstone boson associated with the spontaneous break down of 
the Peccei-Quinn (PQ) symmetry~\cite{Peccei:1977hh,QCD-axion}, 
is an ideal candidate for such light scalars.
As is well known, the QCD axion can explain the observed dark matter density for the decay constant
$f_a = {\cal O}(10^{11-12})$\,GeV, barring fine-tuning of the initial misalignment angle.
The isocurvature bounds on the QCD axion dark matter were extensively studied in the literature 
(see Refs.\cite{Axenides:1983hj,Seckel:1985tj,Turner:1990uz} for early works, Refs.~\cite{Kawasaki:2008sn,Langlois:2008vk,Hikage:2012be}
for non-Gaussianity of isocurvature perturbations, and Refs.~\cite{Higaki:2014ooa,Marsh:2014qoa,Visinelli:2014twa,Choi:2014uaa,Chun:2014xva} 
for the recent works after BICEP2).
The upper bound on the inflation scale is roughly given by $H_{\rm inf} \lesssim 10^{7} {\rm \,GeV} (f_a/\GEV{11})^{0.408}$~\cite{Ade:2013zuv},
which shows clear tension with high-scale inflation.
There have been proposed several solutions to the tension between the high-scale
inflation and the QCD axion dark matter (cf. \cite{Higaki:2014ooa} and references therein).

If there are other light scalars, we can similarly apply the isocurvature constraints. 
A plausible candidate for such light scalars is string axions, the imaginary components of the moduli fields \cite{Svrcek:2006yi}. 
There appear many moduli fields through compactifications in the string theory.
In order to have a sensible low-energy theory, those moduli fields must be stabilized properly.
Many of them can be stabilized with a heavy mass because of the fluxes \cite{Grana:2005jc,Blumenhagen:2006ci}, whereas
some of them  may remain relatively light and play an important cosmological role.
In particular, the axions respect the following axionic shift symmetry,
\bea
a \to a + C,
\label{shift_symmetry}
\eea
where $C$ is a real transformation parameter. The axion acquires a non-zero mass through
 some non-perturbative effects which explicitly break the shift symmetry. Depending on the nature of the shift-symmetry breaking,
the axion mass can be extremely light~\cite{Arvanitaki:2009fg}. Then such light axions become so long-lived that they contribute to dark matter.

In this letter we study a possibility to solve the tension between the string axion dark matter and high-scale inflation
suggested by BICEP2 by a late-time entropy production due to thermal 
inflation \cite{Yamamoto:1985rd,Lazarides:1985ja,Lyth:1995hj,Lyth:1995ka,Asaka:1997rv,Asaka:1999xd}. 
One of the main differences of the string axions from general moduli fields in this context is that no secondary coherent oscillations are
induced at the end of thermal inflation,  because the string axions do not receive the so called Hubble-induced mass.
This greatly helps to dilute light axions efficiently,  as the effect of the secondary coherent oscillations becomes prominent for light moduli fields.
We will show that a large portion of the parameter space can be indeed consistent with the isocurvature bound in the presence
of large entropy production by thermal inflation. We will also study the viability of axion dark matter 
with mass about $7$\,keV~\cite{Higaki:2014zua,Jaeckel:2014qea,Lee:2014xua}\footnote{
See also Refs.~\cite{Kawasaki:1997ah,Hashiba:1997rp,Asaka:1997rv} for the early works on the
X-ray constraint on a light modulus field.
}  as the
origin of the $3.5$\,keV X-ray line excess~\cite{Bulbul:2014sua,Boyarsky:2014jta} in the presence of 
such late-time entropy production.

%%%%%%%%%%%%%%%%%%%%%%%%%%%%%%%%%%%%%%
\section{Isocurvature perturbations of axions}
\label{sec:ipa}
%%%%%%%%%%%%%%%%%%%%%%%%%%%%%%%%%%%%%%

We briefly discuss isocurvature perturbations induced by axion coherent oscillations.
Throughout this letter we assume a large axion decay constant, $f_a$, of order $\GEV{15}$, 
for which thermal production is negligible.  Also neglected is non-thermal production by the saxion
decays~\cite{Cicoli:2012aq,Higaki:2012ar,Higaki:2013lra}. In general both contributions do not
induce isocurvature perturbations.

The potential of the axion is given by
\beq
	V(a) = m_a^2 f_a^2 \bigg[ 1-\cos\bigg( \frac{a}{f_a} \bigg) \bigg],
\eeq
where the potential minimum is located at the origin.  The axion potential
 can be well approximated as the quadratic potential with mass $m_a$ in the vicinity of the minimum.
In linear perturbation theory, the power spectrum of the axion isocurvature perturbation 
can be expressed in terms of the axion fluctuation as follows;
\beq
	\mathcal{P}^{1/2}_{\mathcal{S},a} = \frac{\delta \rho_a}{\rho_a} = \frac{2 f_a \theta_i \delta a + (\delta a)^2}{(f_a \theta_i)^2 + (\delta a)^2}
	\label{axion_isocurv}
\eeq
where axion field fluctuation is given by $\delta a = H_{\rm inf}/2 \pi$, 
$\theta_i$ is an initial misalignment angle of the axion coherent oscillation 
and we assume that the potential of the axion can be well approximated by the quadratic one.
Here and in what follows we focus on a single axion contribution to the CDM isocurvature perturbations, for simplicity. 
The CDM isocurvature perturbation is related to the axion isocurvature perturbation 
by $$\mathcal{P}_{\mathcal{S},{\rm CDM}} = \lrfp{\Omega_a}{\Omega_{\rm CDM}}{2} \mathcal{P}_{\mathcal{S},a},$$
where $\Omega_a$ and $\Omega_{\rm CDM}$ denote the density parameter of the axion and the total CDM, respectively. 
The CDM isocurvature perturbation is tightly constrained by Planck~\cite{Ade:2013zuv};
\beq
	\mathcal{P}_{\mathcal{S},{\rm CDM}} < \frac{\beta}{1-\beta} \mathcal{P}_{\mathcal{R}} ~~~\text{with}~~~ \beta = 0.039 
	\label{iso-Planck}
\eeq
at $95 \%$ CL, and $\mathcal{P_R} \approx 2.2 \times 10^{-9}$ is the curvature perturbation.
This places  a very stringent upper bound on the axion density parameter for high-scale inflation. 

The axion starts to oscillate about the potential minimum when the mass becomes comparable to the Hubble parameter.
Assuming that  it starts to oscillate in the radiation dominated era after  reheating, we can write down
the energy-to-entropy ratio, $\rho_a / s$, as
\beq
	\frac{\rho_a}{s} = \frac{1}{8} T_{\rm osc} \bigg( \frac{a_{\rm osc}}{M_P} \bigg)^2,
	\label{rho_a_over_s}
\eeq
where $a_{\rm osc}$ is the initial amplitude defined as $a_{\rm osc} = {\rm max}(f_a \theta_{\rm osc},~H_{\rm inf}/ 2\pi)$, 
and we have defined $\theta_{\rm osc}^2 = \theta_i^2 f(\theta_i)$. 
Here $f(\theta_i)$ represents the anharmonic effect for 
the axion coherent oscillation~\cite{Lyth:1991ub,Bae:2008ue,Visinelli:2009zm,Kobayashi:2013nva}, and it is unity for $\theta_i \ll \pi$, while it diverges  as $\theta_i$ approaches $\pi$ where the axion potential becomes maximum. 
Also we have defined $T_{\rm osc}$ as the temperature at the beginning of the axion oscillation: 
\beq
	T_{\rm osc} = \bigg( \frac{90}{\pi^2 g_*(T_{\rm osc})} \bigg)^{1/4} \sqrt{m_a M_P},
	\label{T_osc}
\eeq
where $g_*(T)$ is the relativistic degrees of freedom at cosmic temperature $T$,  and % taken to be $228.75$ hereafter and 
$M_P$ is the reduced Planck mass. Here and in what follows, we neglect the thermal effect on the axion mass.
The present density parameter of the axion can then be calculated by using $\rho_{\rm cr,0}/s_0 = 3.64 \times 10^{-9} h^2~{\rm GeV}$ as following;
\beq
	\Omega_a h^2 \simeq
	\begin{cases}
		& 4 \times \bigg(\cfrac{228.75}{g_*(T_{\rm osc})} \bigg)^{1/4} \bigg( \cfrac{m_a}{10^{-9}~{\rm eV}} \bigg)^{1/2}
		\bigg( \cfrac{f_a}{10^{15}~{\rm GeV}} \bigg)^2 \theta_i^2f(\theta_i) ~~~\text{for}~~~ \theta_i > \theta_c \\[4mm]
		& 1 \times 10^{-3} \bigg(\cfrac{228.75}{g_*(T_{\rm osc})} \bigg)^{1/4} \bigg( \cfrac{m_a}{10^{-9}~{\rm eV}} \bigg)^{1/2}
		\bigg( \cfrac{H_{\rm inf}}{10^{14}~{\rm GeV}} \bigg)^2 ~~~\text{for}~~~ \theta_i < \theta_c
	\end{cases}
	\label{Omega_a_noEP}
\eeq
where $\theta_c \approx 0.016(H_{\rm inf}/10^{14}~{\rm GeV})(10^{15}~{\rm GeV}/f_a)$
represents the critical value where the quantum fluctuation $\delta a$ becomes equal to
the classical field deviation $\theta_i f_a$. 
Note that the high-scale inflation implies the axion overproduction 
unless the axion mass is extremely light such as, $m_a \lesssim 10^{-9}~{\rm eV}$, 
for $f_a \sim 10^{15}~{\rm GeV}$.

Let us see how severe is the isocurvature bound (\ref{iso-Planck}) on the axion dark matter.
To this end, let us take  $H_{\rm inf} = 10^{14}$\,GeV, $f_a = \GEV{15}$, and $m_a = 10^{-9}$\,eV as reference
values.   For $\theta_i = 0.1 (>\theta_c)$, the axion density parameter is about $\Omega_{a} h^2 \sim 0.04$.
Then the resultant CDM isocurvature perturbations, ${\cal P}_{{\cal S},{\rm CDM}}$, would exceed the 
bound by about $8$ orders of magnitude. For $\theta_i < \theta_c$, the axion isocurvature perturbations become
highly non-Gaussian, and the CDM isocurvature perturbations exceed the bound by more than $5$ orders of magnitude. 
Thus, one needs some extension in order to resolve the tension between the string axions and the high-scale inflation.
One possibility is to induce huge entropy production after the axion starts to oscillate, which would significantly
reduce the axion abundance, thereby relaxing the tension.  In the next section we consider thermal inflation as 
such late-time entropy production.

%%%%%%%%%%%%%%%%%%%%%%%%%%%%%%%%%%%%%%
\section{Thermal Inflation}
\label{sec:TI}
%%%%%%%%%%%%%%%%%%%%%%%%%%%%%%%%%%%%%%

In this section, we briefly review thermal inflation. Thermal inflation is one of the attractive solutions 
to the overproduction problem of unwanted relics. In order to realize thermal inflation, we introduce a scalar field,
$\phi$,  the so-called flaton, which has a relatively flat potential. 
The flaton is assumed to have couplings with thermal plasma so that it acquires a thermal mass and 
stabilized at the origin for a while.  While it was trapped at the origin,  the vacuum energy of the flaton drives mini-inflation. 
The thermal inflation lasts  until the origin becomes unstable.

The thermal inflation model is often considered in the framework of supersymmetry (SUSY) as it is easy to realize
the required flat potential. Then the flaton lives in a chiral supermultiplet $\phi$, which has the following superpotential,\footnote{
With an abuse of notation, we shall use the same symbol to denote both a chiral superfield and its lowest component.
} 
\beq
	W = \frac{\phi^n}{n M^{n-3}} + k\phi Q \bar{Q} + W_0,
\eeq
where $Q$ and $\bar{Q}$ are additional vector-like quarks having $SU(3)_{\rm QCD}$ charge, 
$M$ is some cutoff energy scale, $k$ is a Yukawa coupling constant, and  
$W_0$ is the constant term related to the gravitino mass by $W_0 \simeq m_{3/2}M_P^2$.
Including the SUSY breaking effect and finite temperature effect, the flaton potential can be expressed as 
\beq
	V = V_0 -(m^2-T^2) |\phi|^2 + (n-3) \bigg( \frac{A \phi^n}{nM^{n-3}} + {\rm h.c.} \bigg) + \frac{|\phi|^{2(n-1)}}{M^{2(n-3)}},
	\label{flaton_pot}
\eeq
where $m$ and $A$ are the soft mass and the A-term, and 
we have omitted a numerical factor of order unity for the thermal mass term. 
The precise values of the soft parameters depend on the SUSY breaking mediation mechanism.
We assume  gravity mediation, for which $m \sim A \sim m_{3/2} \sim $ TeV or heavier. Then, in contrast to
the thermal inflation model considered in Refs.~\cite{Asaka:1997rv,Asaka:1999xd}, it is possible to
kinematically  forbid the flaton decay into a pair of  $R$-axions.

Assuming that the zero temperature vacuum expectation value (VEV) of $\phi$ is determined by the negative 
mass term and  the last term in (\ref{flaton_pot}), we obtain the VEV as
\beq
	v \equiv \langle \phi \rangle = \bigg( \frac{m M^{n-3}}{\sqrt{n-1}} \bigg)^{1/(n-2)}.
\eeq
The flaton mass $m_\phi$ around the minimum is given by $m_\phi = \sqrt{2(n-2)}m$.
Then, $V_0$ is determined by requiring that the vacuum energy should vanish at the potential minimum, which leads to 
\beq
	V_0 = \frac{n-2}{n-1}m^2 v^2.
\eeq
The Hubble parameter during thermal  inflation $H_{\rm TI}$  is given by
\beq
	H_{\rm TI} \simeq  2 ~{\rm keV} \bigg( \frac{n-2}{n-1} \bigg)^{1/2} \bigg( \frac{m}{1~{\rm TeV}} \bigg) \bigg( \frac{v}{10^{10}~{\rm GeV}} \bigg).
\eeq

Before thermal inflation, the universe is considered to be 
dominated by radiation whose energy density evolves like $\rho_r \propto a^{-4}$, 
so $\rho_r$ is overtaken by $V_0$ at some time and then the thermal inflation starts.\footnote{
We assume that the inflaton decays much before thermal inflation starts.
}
Hence, the temperature at the beginning of the thermal inflation is given by
\beq
	T_{\rm TI} = \bigg( \frac{30}{\pi^2 g_*(T_{\rm TI})} \bigg)^{1/4} V_0^{1/4}.
\eeq
Thermal inflation lasts as long as the flaton is stabilized at the origin because of the positive thermal mass, 
 and it ends when the negative SUSY breaking mass dominates over the thermal mass, i.e., $T_{\rm end} = m$.
Then the flaton starts to roll down the potential and oscillate about its VEV. Thus, the universe is dominated
by coherent oscillations of the flaton, until it decays into ordinary matter.

We assume that the flaton has a mass comparable to its axionic partners ($R$-axions) so that its decay into 
$R$-axions is kinematically forbidden. Then, the flaton mainly decays into a pair of gluons through the $Q,\bar{Q}$ loop 	
with the decay rate expressed as
\beq
	\Gamma_\phi \simeq \frac{\alpha_s^2}{64\pi^3} \frac{m_\phi^3}{v^2},
\eeq
where $\alpha_s$ denotes the strong gauge coupling constant.\footnote{%%
	The decay into gluinos can be kinematically forbidden because the flaton mass is comparable to the gaugino masses in gravity mediation. If not, non-thermally produced LSPs might overclose the universe, and one has to introduce either R-parity violation or a very light LSP. This however does not affect the argument in the text. 
} 
The flaton decay temperature is given by
\bea
	T_R &=& \bigg( \frac{90}{\pi^2 g_*} \bigg)^{1/4} \sqrt{\Gamma_\phi M_P} \non \\
	&\simeq& 5~{\rm GeV} \bigg(\frac{\alpha_s}{0.1} \bigg) \bigg( \frac{228.75}{g_*(T_R)} \bigg)^{1/4} 
	\bigg( \frac{m_\phi}{1~{\rm TeV}} \bigg)^{3/2} \bigg( \frac{10^{10}~{\rm GeV}}{v} \bigg).
\eea
where $c$ is some numerical constant. 

The flaton decay produces a  huge amount of entropy, thereby diluting any relics produced before thermal inflation.
The dilution factor, $\Delta$, is defined by the ratio of the the entropy densities before and after the thermal inflation;
\beq
	\begin{split}
	\Delta &\equiv \frac{S_{\rm after}}{S_{\rm before}} = \frac{30}{\pi^2 g_*(T_{\rm end})} \frac{V_0}{T_{\rm end}^3 T_R} \\[1mm]
	&\simeq 1 \times 10^{14} \bigg( \frac{n-2}{n-1} \bigg) \bigg( \frac{228.75}{g_*(T_{\rm end})} \bigg) 
	\bigg( \frac{1~{\rm TeV}}{m} \bigg) \bigg( \frac{10~{\rm GeV}}{T_R} \bigg) \bigg( \frac{v}{10^{10}~{\rm GeV}} \bigg)^2.
	\end{split}
\eeq
Thus the axion abundance can be reduced significantly, which greatly relaxes the isocurvature constraint.
Note that the dilution can be realized only if the axion starts to oscillate before the thermal inflation, 
which requires $m_a > H_{\rm TI}$.

%%%%%%%%%%%%%%%%%%%%%%%%%%%%%%%%%%%%%%
\section{Relaxing the isocurvature bounds on axions}
\label{sec:aa}
%%%%%%%%%%%%%%%%%%%%%%%%%%%%%%%%%%%%%%

The thermal inflation produces a huge amount of entropy, so the isocurvature bound on the axion dark matter
is expected to be relaxed. Here we investigate which parameter space is allowed in the presence of late-time
entropy production due to thermal inflation.
To this end we consider the following three cases in turn:
the axion starts to oscillate 
(A) before the thermal inflation, i.e. $m_a > H_{\rm TI}$, 
(B) after the thermal inflation before the flaton decay, i.e. $H_{\rm TI} > m_a > \Gamma_\phi$, 
and (C) after the flaton decay, i.e. $m_a < \Gamma_\phi$.

First, we consider the case (A). 
Since the axion starts to oscillate in radiation dominated phase, the temperature at the beginning of the axion oscillation is given by (\ref{T_osc}).
Then, taking into account the entropy production due to the thermal inflation, 
the present axion abundance is given by eq.~(\ref{rho_a_over_s}) divided by $\Delta$.
Then, we get the density parameter of the axion in terms of the axion mass as
\beq
	\Omega_a h^2 \simeq 4 \times 10^{-7} \bigg(\frac{228.75}{g_*(T_{\rm osc})} \bigg)^{1/4} \bigg( \frac{m_a}{1~{\rm keV}} \bigg)^{1/2}
	\bigg( \frac{a_{\rm osc}}{10^{15}~{\rm GeV}} \bigg)^2 \bigg( \frac{10^{13}}{\Delta} \bigg)
	% f(\theta_i)
	~~~\text{for}~~~ m_a > H_{\rm TI}.
	\label{Omega_a_1}
\eeq

Let us comment on  the secondary oscillations of the moduli field.
In general,  the moduli field receives the so called Hubble-induced mass term through Planck-suppressed 
interactions with the flaton.
Then, during thermal inflation, the minimum of the moduli potential is slightly displaced from the true minimum.
When the thermal inflation ends,  the potential minimum moves toward the true minimum with a timescale of $m^{-1}$, 
which is much shorter than the moduli oscillation period. 
Therefore, for the moduli fields, the shift of the potential minimum takes place instantly, 
which induces the secondary coherent oscillations~\cite{Lyth:1995ka,Asaka:1997rv,Asaka:1999xd}. 
However, this is not the case for the axions 
because the Hubble-induced mass is forbidden by the shift symmetry (\ref{shift_symmetry}).\footnote{
	Precisely speaking, the axion potential is affected by the flaton potential through the gravitational interaction, 
	and so, the axion mass changes slightly at the end of thermal inflation. 
	This, however,  does not change our discussion because its contribution is negligibly small 
	as shown in Appendix~\ref{app:sec_osc}.
}

Next, we consider the case (B), where
the axion starts to oscillate when the universe is dominated by the flaton coherent oscillations.
The axion abundance is given by
\beq
	\frac{\rho_a}{s} = \frac{1}{8} T_R \bigg( \frac{a_{\rm osc}}{M_P} \bigg)^2, 
\eeq
which leads to 
\beq
	\Omega_a h^2 \simeq 5 \times \bigg( \frac{T_R}{1~{\rm GeV}} \bigg) \bigg( \frac{a_{\rm osc}}{10^{15}~{\rm GeV}} \bigg)^2 
	~~~\text{for}~~~ H_{\rm TI} > m_a > \Gamma_\phi.
	\label{Omega_a_2}
\eeq
The case (C) is nothing but the scenario without thermal inflation and the resultant density parameter is given by (\ref{Omega_a_noEP}).

\begin{figure}[H]
\centering
\subfigure[~$\Delta = 10^{10}$]{
\includegraphics [width = 7.5cm, clip]{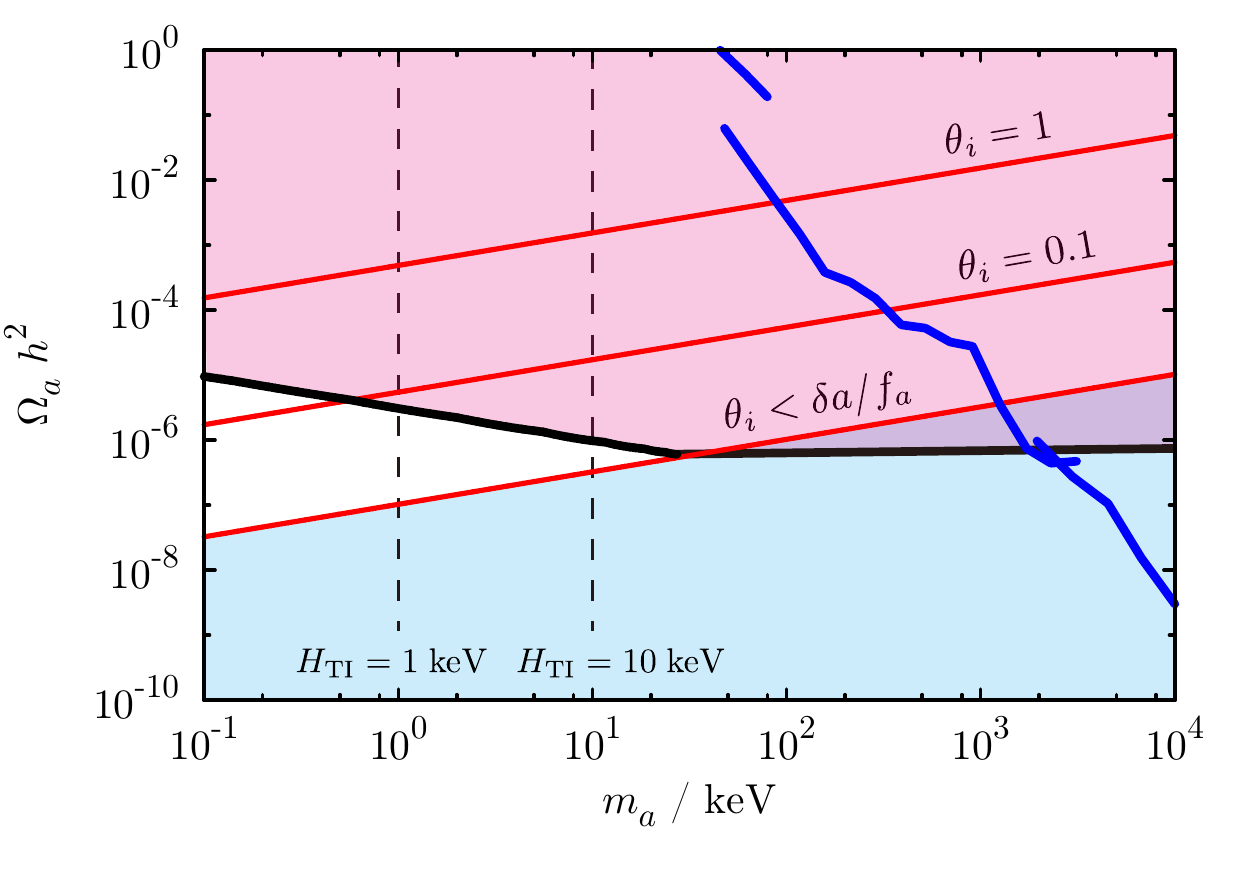}
\label{contour_1}
}
\subfigure[~$\Delta = 10^{12}$]{
\includegraphics [width = 7.5cm, clip]{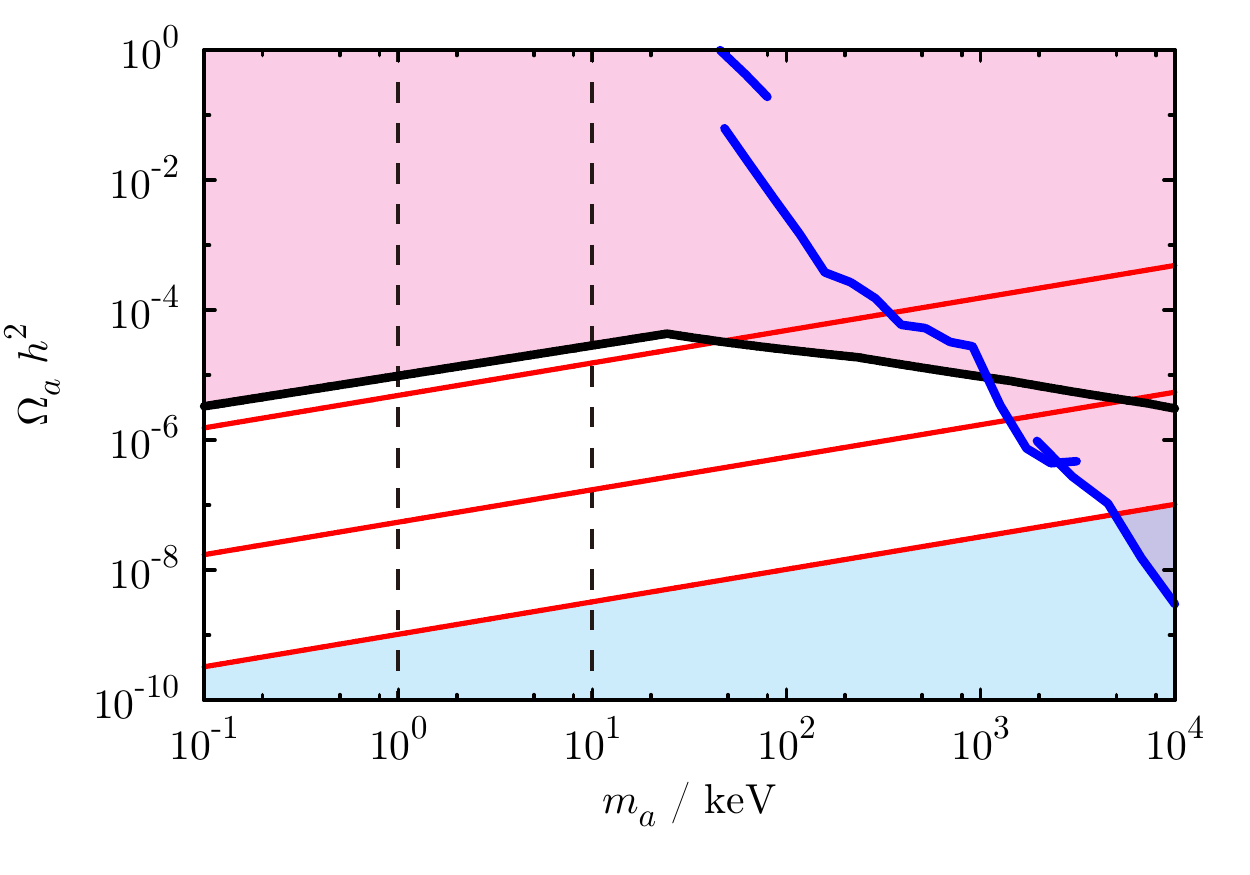}
\label{contour_2}
}
\subfigure[~$\Delta = 10^{14}$]{
\includegraphics [width = 7.5cm, clip]{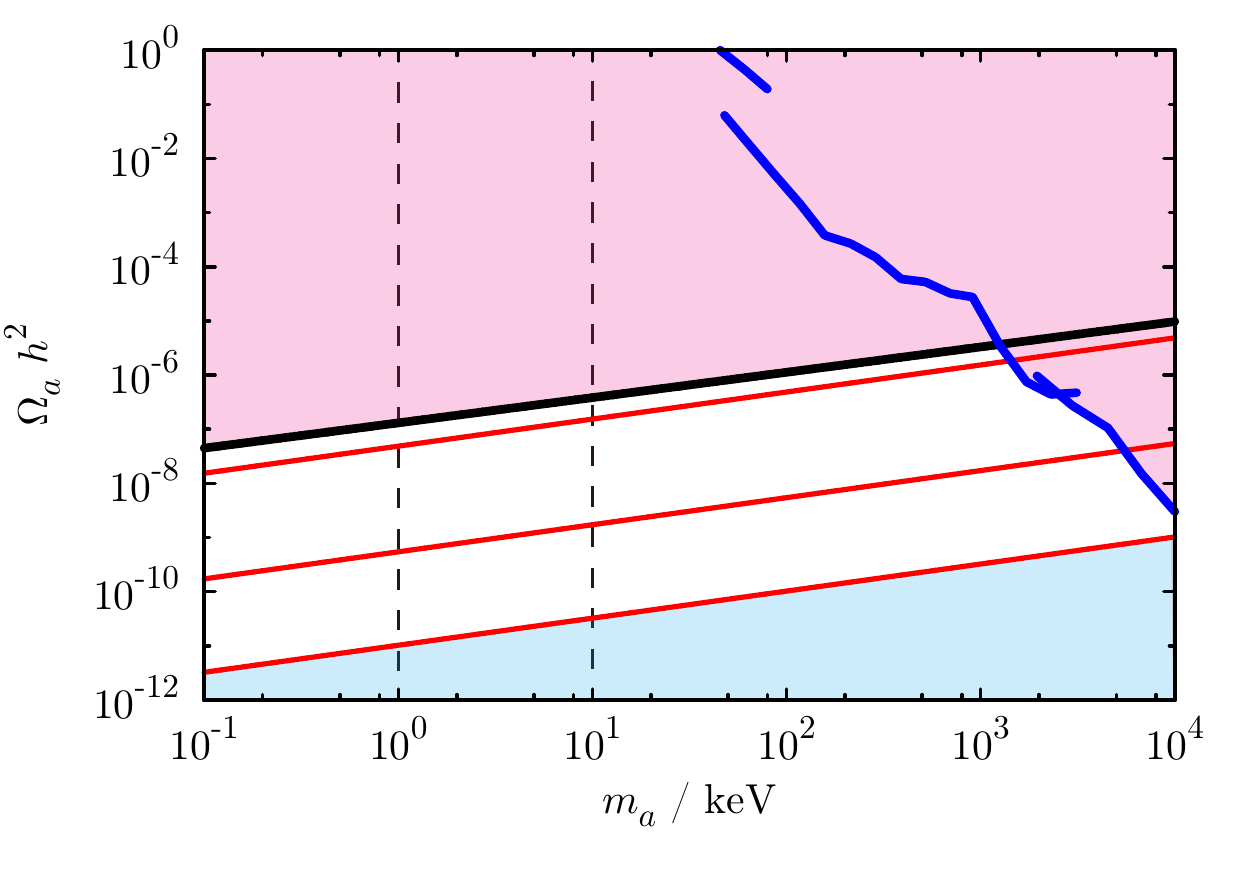}
\label{contour_3}
}

\caption{
	Constraints on the axion parameters on $\Omega_ah^2$--$m_a$ plane. The upper shaded (magenta) region
	is excluded either by the isocurvature perturbation (black line) or by the X-ray observations (blue line).
	The lower shaded (cyan) region cannot be realized because of the absolute lower limit on the axion abundance due to the quantum fluctuation, $\delta a$.
		The dashed vertical lines represent the typical Hubble parameter during inflation, 
		$H_{\rm TI} = 1~{\rm keV}$ (left) and $10~{\rm keV}$ (right). For the successful dilution, the axion mass must be heavier than $H_{\rm TI}$. 
	We have taken $f_a = 10^{15}~{\rm GeV}$, $H_{\rm inf} = 10^{14}~{\rm GeV}$ 
	and $\Delta = 10^{10}$, $10^{12}$ and $10^{14}$ in Fig.~\ref{contour_1}, \ref{contour_2} and \ref{contour_3} respectively.
}
\label{contour}
\end{figure}

Using Eqs.~(\ref{Omega_a_1}),(\ref{Omega_a_2}),(\ref{Omega_a_noEP}) and (\ref{axion_isocurv}), 
we estimated the isocurvature bounds in the $\Omega_a$--$m_a$ plane as shown in Fig.~\ref{contour}.
Furthermore we also take account of the bounds from diffuse X($\gamma$)-ray observations 
such as HEAO-1 \cite{Gruber:1999yr}, INTEGRAL \cite{Bouchet:2008rp}, COMPTEL \cite{comptel} 
because the axion can generically decay into two photons via
\bea
{\cal L}_{a \gamma \gamma } = \frac{\alpha_{\rm EM}}{4\pi}
\frac{a}{f_a} F_{\mu\nu} \tilde F^{\mu\nu} = \frac{g_{a \gamma \gamma}}{4}  a F_{\mu\nu} \tilde F^{\mu\nu},
\eea
where $\alpha_{\rm EM}$ is the fine-structure constant. For $f_a \sim \GEV{15}$, the axion-photon
coupling constant $g_{a \gamma \gamma}$ is about $1/M_P$. 
  The decay rate of the axion to two photons is given by
\bea
\Gamma(a \to \gamma \gamma) = \frac{g_{a \gamma \gamma}^2}{64 \pi} m_a^3.
\eea
In Fig.~\ref{contour}, there are three panels corresponding to the dilution factor $\Delta = 10^{10}, 10^{12}$, and 
$10^{14}$, where we have fixed $H_{\rm inf} = \GEV{14}$ and $f_a = \GEV{15}$.
In each panel, the upper shaded (magenta) region is excluded by
the bounds from the isocurvature perturbation (black) and X-ray observation (blue). 
The diagonal (red) lines correspond to $\theta_i = 1$ (upper), $\theta_i = 0.1$ (middle) and $\theta_i \leq \theta_c$  (lower), respectively.  
Note that the shaded (cyan) region below the lower diagonal line 
cannot be realized because a certain amount
of axions is necessarily produced by the quantum fluctuations of the axion field, thus there is an absolute lower bound
on the axion abundance. Note that we have taken account of the anharmonic effects on the axion abundance.
The vertical dashed lines show the typical Hubble parameter during thermal inflation, $H_{\rm TI} = 1$\,keV and $10$\,keV. 
The region with $m_a < H_{\rm TI}$ is excluded as the axion abundance is not diluted
by thermal inflation. That is to say, if $H_{\rm TI} = 1$\,keV, the region left to the left vertical dashed line is excluded. 
The allowed region is the white region surrounded by the shaded regions and the vertical line given by 
$m_a = H_{\rm TI}$. From the figure one can see that the allowed region is extended to heavier axion masses for 
a larger entropy dilution factor. The axion, however, cannot account for the total CDM density, for 
$H_{\rm inf} = \GEV{14}$ and $f_a = \GEV{15}$.

\begin{figure}[H]
\centering
\subfigure[~$H_{\rm inf} = 10^{14}~{\rm GeV}$]{
\includegraphics [width = 7.5cm, clip]{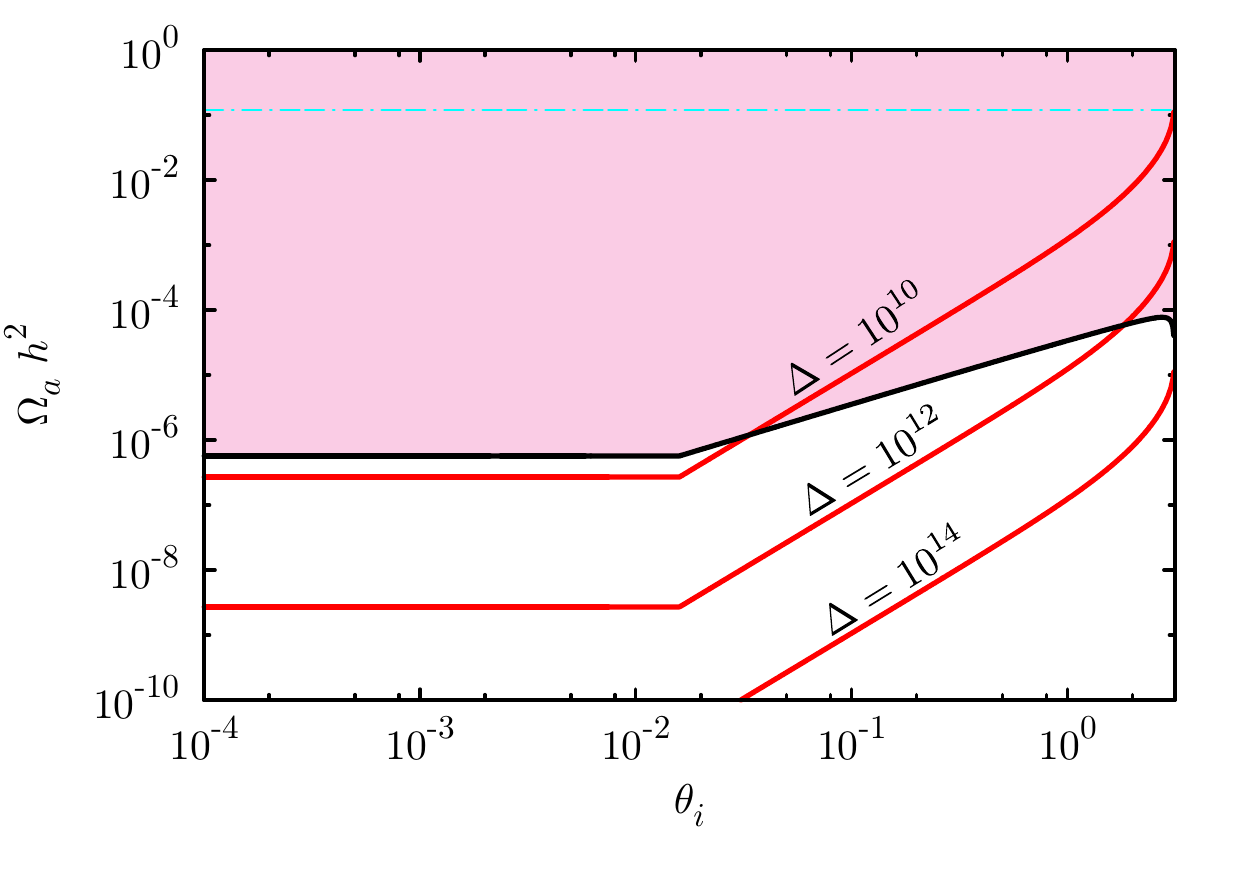}
\label{7keV_1}
}
\subfigure[~$H_{\rm inf} = 10^{12}~{\rm GeV}$]{
\includegraphics [width = 7.5cm, clip]{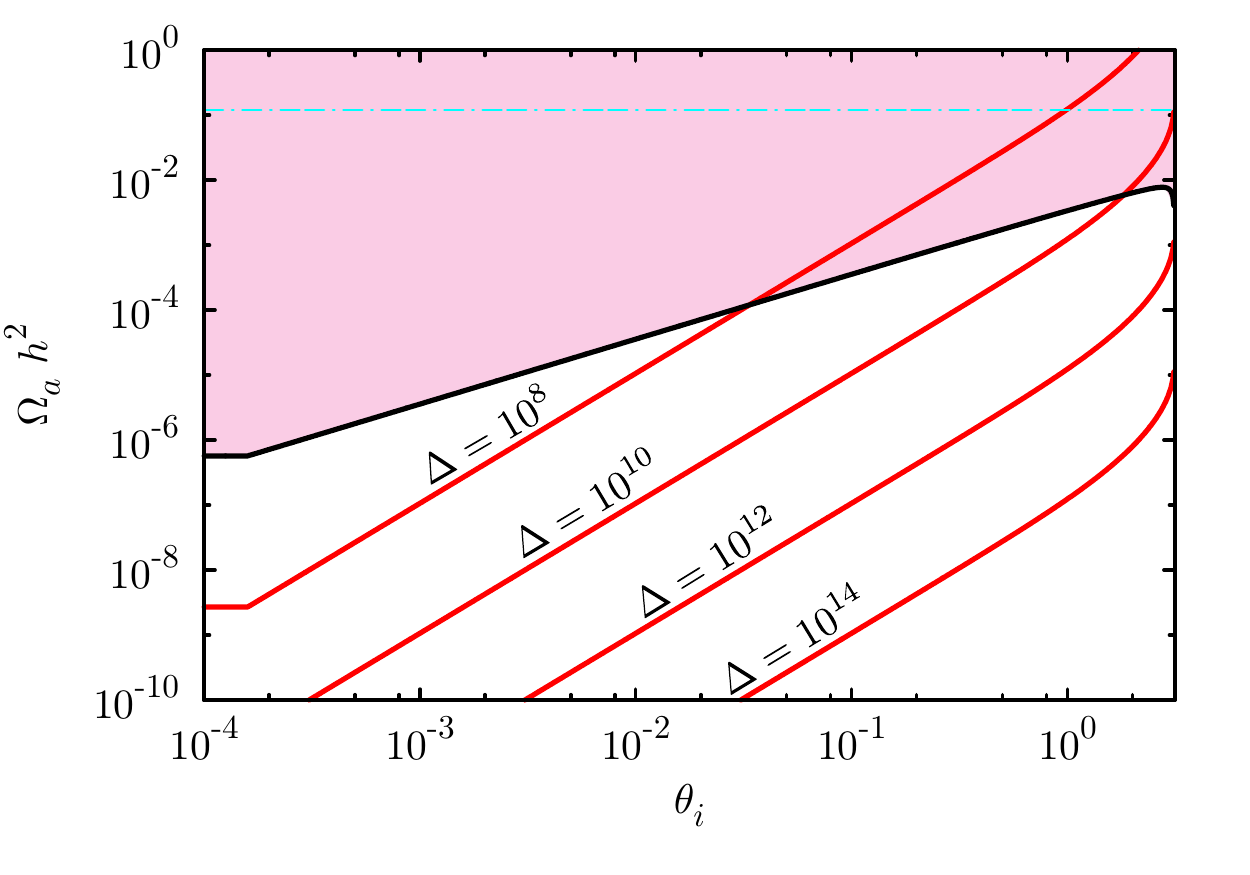}
\label{7keV_2}
}
\subfigure[~$H_{\rm inf} = 10^{10}~{\rm GeV}$]{
\includegraphics [width = 7.5cm, clip]{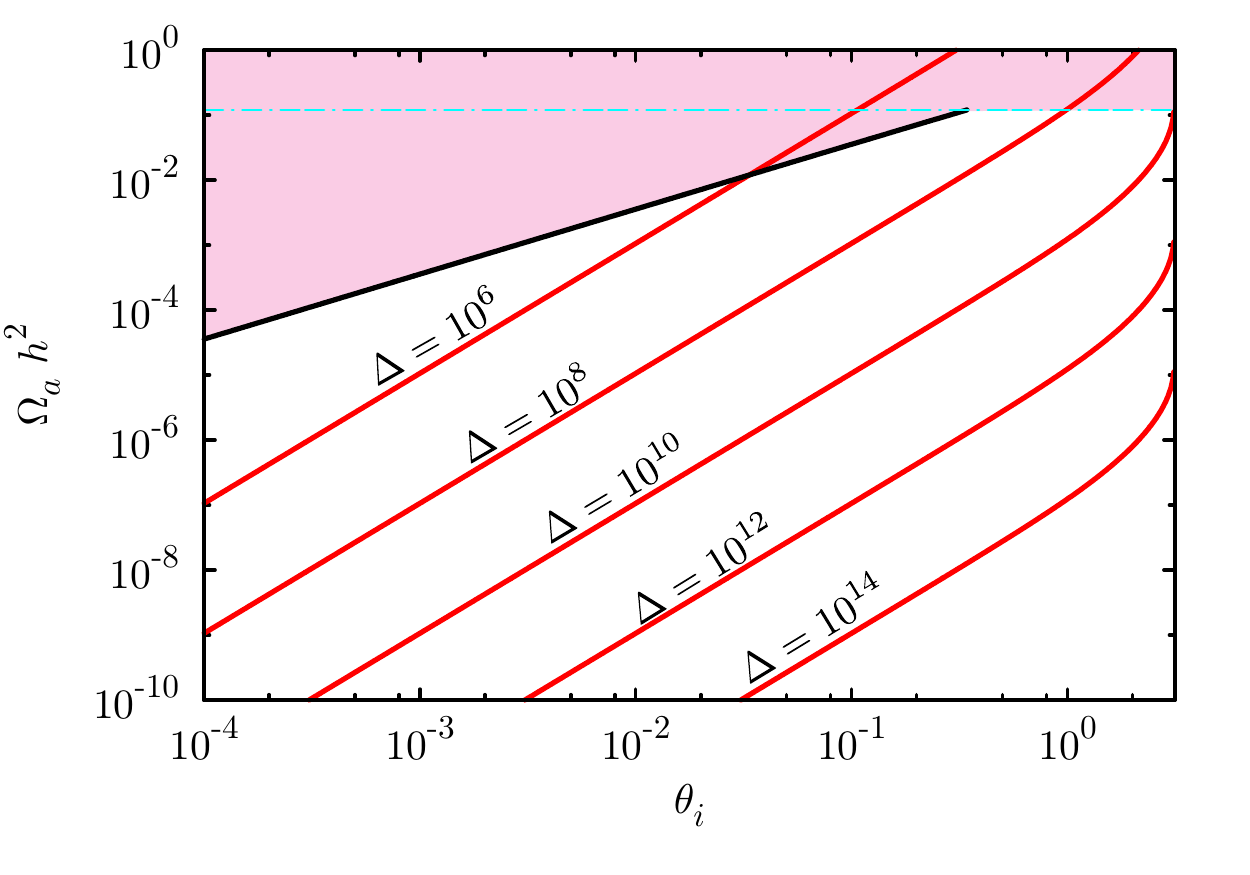}
\label{7keV_3}
}

\caption{
	Constraints from the isocurvature perturbation for the axion with $m_a = 7~{\rm keV}$ 
	on the $\Omega$--$\theta_i$.
	The solid (red) curves show the axion abundance as a function of the initial misalignment angle $\theta_i$ 
	for various values of the dilution factor. The upper shaded (magenta) region is excluded.
	We have taken $f_a = 10^{15}~{\rm GeV}$ 
	and $H_{\rm inf} = 10^{14}~{\rm GeV}$, $10^{12}~{\rm GeV}$ and $10^{10}~{\rm GeV}$ 
	in Fig.~\ref{7keV_1}, \ref{7keV_2} and \ref{7keV_3} respectively.
}
\label{fig:7keV}
\end{figure}

Especially focusing on the axion with $m_a = 7~{\rm keV}$, we show
the constraint from the isocurvature perturbation  in Fig.~\ref{fig:7keV}, where we have varied the initial misalignment angle $\theta_i$. Such $7$\,keV axion dark matter decaying into
two photons can be a possible origin of the recently found the $3.5$\,keV X-ray line 
excess~\cite{Higaki:2014zua,Jaeckel:2014qea,Lee:2014xua}. In Fig.~\ref{fig:7keV}, we show three panels corresponding
to $H_{\rm inf} = \GEV{14}, \GEV{12}$ and $\GEV{10}$.
From Fig.~\ref{fig:7keV}, one can see that the 7 keV axion cannot be the dominant component of the present CDM 
in high-scale inflation with $H_{\rm inf} = \GEV{14}$, whereas  it is possible for $H_{\rm inf} \lesssim 10^{10}~{\rm GeV}$.
In the case of $H_{\rm inf} \simeq 10^{10}~{\rm GeV}$, at least $\Delta \sim 10^7$ is required for the axion 
to account for the total CDM.

%%%%%%%%%%%%% MULTI-FIGURE  %%%%%%%%%%%%% 
%%%%%%%%%%%%%%%%%%%%%%%%%%%%%%%%%%%

%%%%%%%%%%%%% MULTI-FIGURE  %%%%%%%%%%%%% 
%%%%%%%%%%%%%%%%%%%%%%%%%%%%%%%%%%%

%%%%%%%%%%%%%%%%%%%%%%%%%%%%%%%%%%%%%%
\section{Conclusions}
\label{sec:conc}
%%%%%%%%%%%%%%%%%%%%%%%%%%%%%%%%%%%%%%

If the inflation scale is high, e.g. $H_{\rm inf} \sim \GEV{14}$ as suggested by the BICEP2 result, 
any light scalars acquire  large quantum fluctuations during inflation, which give rise to the CDM 
isocurvature perturbation if they contribute to the present CDM. 
The Planck data allows only a small admixture of isocurvature perturbations,
which severely constrains  the present density parameter of the light scalars. 
One possibility to relax the tension between high-scale inflation 
and the isocurvature bound on the light scalars  is to generate a huge amount of entropy at a later time.

In this letter we have focused on the string axions  with a large decay constant $f_a \sim \GEV{15}$ 
as such a light scalar, and studied its cosmology in the presence of late-time entropy production, 
taking account of the anharmonic effect.
As a concrete example of late-time entropy production, we have considered the thermal inflation, 
and found that the axion coherent oscillations can be efficiently diluted, partly because effectively no secondary coherent
oscillations are induced at the end of inflation, in contrast to a generic modulus field.  
We have shown that there appears a wide parameter region where 
the isocurvature bound  as well as the X-ray observation bound are satisfied. However, the axion 
cannot be the dominant CDM component for $H_{\rm inf} = \GEV{14}$ and $f_a = \GEV{15}$. 
 Also we have studied the viability of the $7$\,keV axion dark matter 
 for various values of the inflation scale, and found that the $7$\,keV axion can be the dominant
component of dark matter if $H_{\rm inf} \lesssim \GEV{10}$ and $\Delta \gtrsim 10^7$.

%%%%%%%%%%%%%%%%%%%%%%%%%%%%%%%%%%%%%
\section*{Acknowledgment}
%%%%%%%%%%%%%%%%%%%%%%%%%%%%%%%%%%%%%
This work was supported by the Grant-in-Aid for Scientific Research on
Innovative Areas (No.23104008 [FT]),  JSPS Grant-in-Aid for
Young Scientists (B) (No.24740135) [FT], Scientific Research (B) (No.26287039 [FT]), 
Scientific Research (C) (No.25400248[MK]), 
Inoue Foundation for Science [FT].  This work was also
supported by World Premier International Center Initiative (WPI Program), MEXT, Japan [MK and FT].

\appendix

\section{Secondary oscillation of the axion}\label{app:sec_osc}
 
Here, we consider a specific axion model and quantitatively calculate the secondary oscillations of the axion.
Focusing only on the axion sector, the K\"ahler potential and superpotential are respectively given by
\beq
	K = \frac{1}{2}(\mathcal{A} + \mathcal{A}^\dag)^2
\eeq
and
\beq
	W = \Lambda^3 e^{-\sqrt{2}\mathcal{A}/f_a} + W_0 + \frac{\phi^n}{n M^{n-3}},
\eeq
where $\mathcal{A}$ is the axion supermultiplet whose scalar component include the saxion ($\sigma$)
and axion denoted as $\mathcal{A} = (\sigma + i a)/\sqrt{2}$ 
and $\Lambda$ is some energy scale corresponding to the explicit symmetry breaking of the
shift symmetry of the axion. We have added the interaction of the flaton, which contributes to  the
VEV of the superpotential after thermal inflation.

The $F$-term scalar potential in supergravity is given by
\beq
	V_F = e^{K/M_P^2} [ K^{ij^*}D_{i}W D_{l^*}W^* - 3|W|^2/M_P^2], 
	\label{F-term}
\eeq
where $D_i W = \partial_i W + (\partial_i K) W /M_P^2$ and $K^{ij^*}$ is the inverse matrix of $K_{ij^*}$.
The axion potential term comes from the second term in the square bracket in (\ref{F-term}) and written as
\beq
	V_a  \simeq - 6 m_{3/2} \Lambda^3 \bigg[ \cos \bigg( \frac{a}{f_a} \bigg) 
	+ \frac{|\phi|^n}{n M^{n-3}m_{3/2}M_P^2} \cos\bigg( \frac{a}{f_a}-n\theta_\phi\bigg)\bigg] + \cdots,
	\label{axion_pot}
\eeq
where $\theta_\phi$ is the argument of $\phi$, the dots represent higher order terms, 
and the saxion is assumed to be stabilized at the origin.
This is indeed the case if $\Lambda^3 \ll m_{3/2} f_a^2$, as the saxion will be much heavier than the axion.
During the thermal inflation, $\phi$ is stabilized at the origin and the second term in square bracket in (\ref{axion_pot}) vanishes.
After the thermal inflation, $\phi$ acquires the large VEV and the second term in square bracket 
becomes non-zero with a time scale of $m^{-1}$, which induces the secondary oscillations. 
The shift of the potential minimum, however,  is suppressed by a factor of $\frac{m_a^2}{m m_{3/2}}$, 
compared to the case of the moduli field with the Hubble-induced mass.  Thus,  
%
%However, this contribution is negligibly small compared to that induced by the so-called Hubble-induced mass
%term, which usual moduli field acquires during thermal inflation. 
%the second term by a factor of $(v/M_P)^2$, 
 the secondary oscillation contribution is negligible in our calculation for the relic abundance of the axion.

\end{document}